# Dark counts of superconducting nanowire single-photon detector under illumination

Sijing Chen, Lixing You,* Weijun Zhang, Xiaoyan Yang, Hao Li, Lu Zhang, Zhen Wang, and Xiaoming Xie

*State Key Laboratory of Functional Materials for Informatics, Shanghai Institute of Microsystem and Information Technology (SIMIT), Chinese Academy of Sciences, 865 Changning Rd., Shanghai 200050, China*

*lxyou@mail.sim.ac.cn*

**Abstract:** An abnormal increase in the SDE was observed for superconducting nanowire single-photon detectors (SNSPDs) when the bias current ($I_b$) was close to the switching current ($I_{sw}$). By introducing the time-correlated single-photon counting technique, we investigated the temporal histogram of the detection counts of an SNSPD under illumination. The temporal information helps us to distinguish photon counts from dark counts in the time domain. In this manner, the dark count rate (DCR) under illumination and the accurate SDE can be determined. The DCR under moderate illumination may be significantly larger than the conventional DCR measured without illumination under a high $I_b$, which causes the abnormal increase in the SDE. The increased DCR may be explained by the suppression of $I_{sw}$ under illumination.

**References and links**

1. G. N. Gol'tsman, O. Okunev, G. Chulkova, A. Lipatov, A. Semenov, K. Smirnov, B. Voronov, A. Dzardanov, C. Williams, and R. Sobolewski, "Picosecond superconducting single-photon optical detector," Appl. Phys. Lett. **79** (6), 705 (2001).
2. C. M. Natarajan, M. G. Tanner, and R. H. Hadfield, "Superconducting nanowire single-photon detectors: physics and applications," Supercond. Sci. Technol. **25** (6), 063001 (2012).
3. M. Sasaki, M. Fujiwara, H. Ishizuka, W. Klaus, K. Wakui, M. Takeoka, S. Miki, T. Yamashita, Z. Wang, A. Tanaka, K. Yoshino, Y. Nambu, S. Takahashi, A. Tajima, A. Tomita, T. Domeki, T. Hasegawa, Y. Sakai, H. Kobayashi, T. Asai, K. Shimizu, T. Tokura, T. Tsurumaru, M. Matsui, T. Honjo, K. Tamaki, H. Takesue, Y. Tokura, J. F. Dynes, A. R. Dixon, A. W. Sharpe, Z. L. Yuan, A. J. Shields, S. Uchikoga, M. Legre, S. Robyr, P. Trinkler, L. Monat, J. B. Page, G. Ribordy, A. Poppe, A. Allacher, O. Maurhart, T. Langer, M. Peev, and A. Zeilinger, "Field test of quantum key distribution in the Tokyo QKD Network," Opt. Express **19** (11), 10387-10409 (2011).

4. L. Van Zwieten, B. P. Singh, S. W. L. Kimber, D. V. Murphy, L. M. Macdonald, J. Rust, and S. Morris, "An incubation study investigating the mechanisms that impact N2O flux from soil following biochar application," Agr Ecosyst Environ **191**, 53-62 (2014).

5. D. M. Boroson, B. S. Robinson, D. V. Murphy, D. A. Burianek, F. Khatri, J. M. Kovalik, Z. Sodnik, and D. M. Cornwell, "Overview and results of the lunar laser communication demonstration," Free-Space Laser Communication and Atmospheric Propagation Xxvi **8971** (2014).

6. V. Burenkov, H. Xu, B. Qi, R. H. Hadfield, and H.-K. Lo, "Investigations of afterpulsing and detection efficiency recovery in superconducting nanowire single-photon detectors," Journal of Applied Physics **113** (21), 213102 (2013).

7. R. Ikuta, H. Kato, Y. Kusaka, S. Miki, T. Yamashita, H. Terai, M. Fujiwara, T. Yamamoto, M. Koashi, M. Sasaki, Z. Wang, and N. Imoto, "High-fidelity conversion of photonic quantum information to telecommunication wavelength with superconducting single-photon detectors," Phys. Rev. A **87** (1), 010301 (2013).

8. T. Yamashita, S. Miki, H. Terai, and Z. Wang, "Low-filling-factor superconducting single photon detector with high system detection efficiency," Opt. Express **21** (22), 27177-27184 (2013).

9. D. Rosenberg, A. J. Kerman, R. J. Molnar, and E. A. Dauler, "High-speed and high-efficiency superconducting nanowire single photon detector array," Opt. Express **21** (2), 1440-1447 (2013).

10. F. Marsili, V. B. Verma, J. A. Stern, S. Harrington, A. E. Lita, T. Gerrits, I. Vayshenker, B. Baek, M. D. Shaw, R. P. Mirin, and S. W. Nam, "Detecting single infrared photons with 93% system efficiency-supplymentary information," Nat. Photonics **7** (3), 210-214 (2013).

11. A. Engel, A. Semenov, H. W. Hübers, K. Ilin, and M. Siegel, "Dark counts of a superconducting single-photon detector," Nucl. Instrum. Methods Phys. Res., Sect. A **520** (1-3), 32-35 (2004).

12. T. Yamashita, S. Miki, W. Qiu, M. Fujiwara, M. Sasaki, and Z. Wang, "Temperature dependent performances of superconducting nanowire single-photon detectors in an ultralow-temperature region," Appl Phys Express **3** (10), 102502 (2010).

13. S. Miki, T. Yamashita, H. Terai, and Z. Wang, "High performance fiber-coupled NbTiN superconducting nanowire single photon detectors with Gifford-McMahon cryocooler," Opt. Express **21** (8), 10208-10214 (2013).

14. F. Marsili, V. B. Verma, J. A. Stern, S. Harrington, A. E. Lita, T. Gerrits, I. Vayshenker, B. Baek, M. D. Shaw, R. P. Mirin, and S. W. Nam, "Detecting single infrared photons with 93% system efficiency," Nat. Photonics **7** (3), 210-214 (2013).

15. E. A. Dauler, M. E. Grein, A. J. Kerman, F. Marsili, S. Miki, S. W. Nam, M. D. Shaw, H. Terai, V. B. Verma, and T. Yamashita, "Review of superconducting nanowire single-photon detector system design options and demonstrated performance," Opt. Eng. **53** (8) (2014).

16. D. M. Boroson, J. J. Scozzafava, D. V. Murphy, B. S. Robinson, and M. I. T. Lincoln, "The lunar laser communications demonstration (LLCD)," 23-28 (2009).

17. A. J. Kerman, D. Rosenberg, R. J. Molnar, and E. A. Dauler, "Readout of superconducting nanowire single-photon detectors at high count rates," J. Appl. Phys. **113** (14), 144511 (2013).

18. M. Fujiwara, A. Tanaka, S. Takahashi, K. Yoshino, Y. Nambu, A. Tajima, S. Miki, T. Yamashita, Z. Wang, A. Tomita, and M. Sasaki, "Afterpulse-like phenomenon of superconducting single photon detector in high speed quantum key distribution system," Opt. Express **19** (20), 19562-19571 (2011).
19. F. Mattioli, R. Leoni, A. Gaggero, M. G. Castellano, P. Carelli, F. Marsili, and A. Fiore, "Electrical characterization of superconducting single-photon detectors," J. Appl. Phys. **101** (5), 054302 (2007).
20. T. Yamashita, S. Miki, K. Makise, W. Qiu, H. Terai, M. Fujiwara, M. Sasaki, and Z. Wang, "Origin of intrinsic dark count in superconducting nanowire single-photon detectors," Appl. Phys. Lett. **99** (16), 161105 (2011).
21. H. Shibata, K. Shimizu, H. Takesue, and Y. Tokura, "Superconducting nanowire single-photon detector with ultralow dark Count rate using cold optical filters," Appl Phys Express **6** (7), 072801 (2013).
22. Q. Zhao, A. McCaughan, F. Bellei, F. Najafi, D. De Fazio, A. Dane, Y. Ivry, and K. K. Berggren, "Superconducting-nanowire single-photon-detector linear array," Appl. Phys. Lett. **103** (14), 142602 (2013).

**1. Introduction**

Superconducting nanowire single-photon detectors (SNSPDs) [1] have been widely accepted as novel near-infrared detectors with high system detection efficiency (SDE), high counting rate, low dark count rate (DCR), and small timing jitter [2]. The merits of SNPDs have been demonstrated in various experiments and applications such as quantum key distribution (QKD) [3-5], free-space laser communication [6], and quantum optics [7].

SDE and DCR are the key parameters of SNSPDs and both are related to the bias current ($I_b$) of the SNSPD. The performance of an SNSPD is often represented as the SDE at a specific DCR [8-10] (e.g., SDE = 50% at DCR = 100 Hz). To measure the SDE and DCR at a given $I_b$, a photon counter is often adopted to count the response pulse with amplitude larger than a specific threshold. The DCR is equivalent to the counts of the photon counter per second ($n_{cb}$) when the SNSPD is blocked from illumination [11, 12]. For measuring the SDE, we should illuminate the SNSPD with a given photon flux ($n_{ph}$) using a pulsed or continuous-wave (CW) light source; then, SDE = $(n_{cp} - n_{cb})/n_{ph}$, where $n_{cp}$ denotes the counts of the photon counter per second when the SNSPD is under illumination. The above definition is widely used in evaluating an SNSPD's performance. In fact, the photon counter cannot distinguish between a photon count and a dark count. Therefore, if $n_{cb}$ measured under illumination is different from that without illumination, the calculation of the SDE may be inaccurate.

Using optical cavity structures, we recently fabricated an NbN SNSPD with an SDE over 75% at DCR = 100 Hz, which is among the best results reported for an SNSPD operated at ~ 2.0 K with similar structures [8, 13-15]. We observed an abnormal increase in the SDE instead of saturation when $I_b$ was close to the switching current ($I_{sw}$). By introducing the time-correlated single-photon counting (TCSPC) technique, we distinguished photon counts from dark counts in the time domain. Therefore, we obtain the DCR of an SNSPD under illumination (DCR$^*$), which is significantly larger than the conventional DCR measured without illumination under the same high $I_b$ even when nph is as low as $10^5$ photons per second (pps). Hence, the corresponding SDE calculated using SDE = $(n_{cp} − n_{cb})/n_{ph}$ is also inaccurate, which leads to the abnormal increase of the SDE.

## 2. Fabrication and characterization of the SNSPD

The substrate used for the SNSPD is a double-sided, thermally oxidized Si wafer with 258-nm-thick $SiO_2$ layers on both sides. The $SiO_2$ layers on the back and front sides of the substrate serve as an anti-reflection layer and a dielectric material of the cavity structure for 1550 nm wavelength, respectively. A 7-nm-thick NbN film was deposited on the Si wafer using DC reactive magnetron sputtering. The films were then patterned into a 70-nm-wide and 130-nm-spaced meandered nanowire covering a square area of 15 μm × 15 μm using e-beam lithography and reactive-ion etching. A 200-nm-thick SiO layer and a 150-nm-thick Ag mirror were deposited on top of the film, which served as the dielectric material and the mirror for the top cavity structure. The thickness of the $SiO_2$ and SiO layers was designed to be 1/4 of the wavelength in order to obtain the highest absorptance for the nanowire at 1550 nm wavelength. Figure 1(a) shows a scanning electron microscope image of the active area of the SNSPD with the optical cavity at the top.

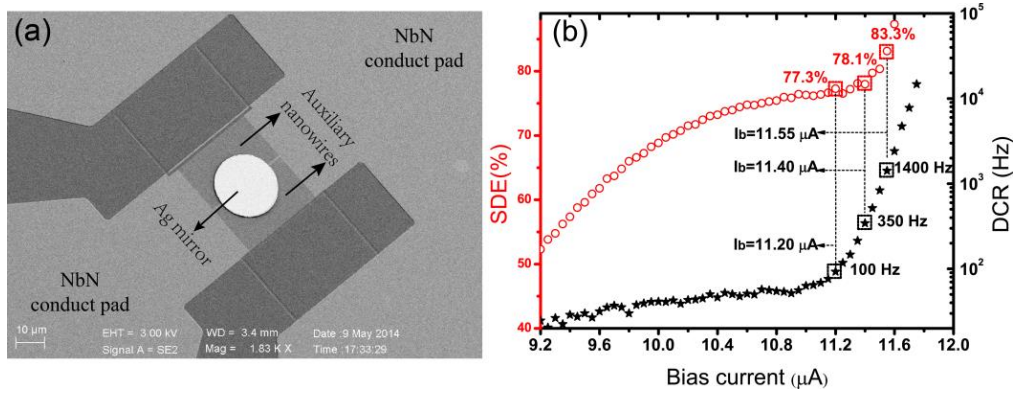

Fig. 1. (a) Scanning electron microscope image of SNSPD tilted at an angle of approximately 20°, the detector meander under the Ag mirror cannot be seen directly. (b) SDE (red circles) and DCR (black stars) as functions of $I_b$, measured at 2.3 K. The SDE and DCR values are marked at three different bias currents: 11.20 μA ($0.95I_{sw0}$), 11.40 μA ($0.97I_{sw0}$), and 11.55 μA ($0.98I_{sw0}$).

For accurately measuring the SDE, $n_{ph}$ is usually set to at least one order of magnitude higher than the maximum DCR (on the order of 10 kHz). The common value is $10^5$ pps or higher. We used a pulsed laser (1550 nm, C10196; Hamamatsu Inc.) with a typical pulse width of 70 ps and various pulse repetition rates ($R_p$). In our experiment, $R_p$ was set to 10 MHz; thus, $n_{ph}$ of $10^5$ pps corresponds to an averaged 0.01 photon per pulse. Figure 1(b) shows SDE and DCR as functions of $I_b$. The $I_{sw}$ of the SNSPD without illumination ($I_{sw0}$) is 11.75 μA, and

SDE is 77.3% when DCR is 100 Hz at 0.95 $I_{sw0}$. The total relative error of the SDE is estimated to be ± 2.5 %, which comes from the uncertainty in power calibration (± 2.5%), attenuator calibration (±0.3%), and the fluctuation of light power in long term (±0.3%). Then the corresponding absolute error of SDE (77%) is 2%. The abnormal phenomenon observed was that the SDE started to increase abruptly rather than saturating when $I_b$ was close to $I_{sw0}$, which is difficult to explain because the contribution of the DCR was already deducted. The highest SDE was 87.3%. Possible causes such as an overshoot of the device current cannot explain these results because both $n_{ph}$ and the counting rate in this experiment are very small [6, 16-18]. One reason could be that the DCR increases when the SNSPD is under illumination. However, the photon counter cannot distinguish between a photon count and a dark count. The key difference between these two types of counts is that dark counts always occur randomly in the time domain, whereas photon counts occur at a specific timeslot for a pulsed laser illumination. If we analyze the temporal histogram for all counts of the SNSPD under illumination, we can distinguish photon counts from dark counts and thus obtain an accurate SDE and DCR. In fact, a TCSPC module may fulfill this function.

## 3. Experiment setup

The system schematics are shown in Fig. 2. To analyze the temporal histogram for the counts of the SNSPD, a TCSPC module (PicoHarp 300) is adopted. Laser pulses with $R_p$ = 10 MHz are heavily attenuated by two variable attenuators with a maximum attenuation of 60 dB each for achieving $n_{ph}$ from $10^5$ to $10^6$ pps. Because the PicoHarp 300 requires negative input signals, the synchronization signal of the laser is inverted, and the output pulse signal from the SNSPD also has a negative pulse using a negative current bias. The inverted synchronized signal is sent to channel "0" of the TCSPC module as a start signal. The output signal from the SNSPD is amplified using a 50 dB low noise amplifier (LNA-650; RF Bay Inc.) and then divided by a 1:1 power divider. One part of the signal is sent to channel "1" of the TCSPC module as the stop signal for temporal analysis, while the other part is sent to a gated photon counter (SR400) for conventional counting, which also provides a reference for the TCSPC module. The TCSPC module not only counts the pulses from the SNSPD, similar to the gated photon counter, but also provides a temporal histogram of all counts.

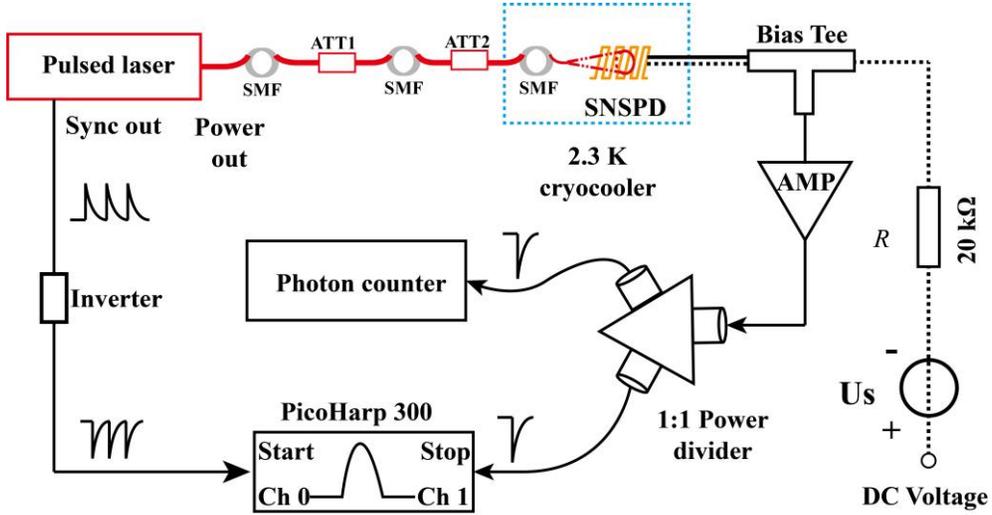

Fig. 2. System schematics for temporal analysis of the output counts of SNSPD. SMF: single-mode fiber; ATT: variable attenuator; AMP: amplifier. The black dashed lines, solid lines with arrows, and red bold lines represent the DC path, RF path, and light path, respectively. The parallel solid and dashed lines represent the path for both the DC and RF signals.

## 4. Temporal analysis of photon counts and dark counts

For accurately counting the response pulse, the waveform of the pulse should be known beforehand. Figure 3(a) shows the response pulse averaged over 200 samples at a bias current of $0.95I_{sw0}$. The inset of Fig. 3(a) shows the electronic noise floor screen-copied by the oscilloscope (DSA 71254, Tektronix) with the peak-to-peak value of 30 mV. The discriminating level of the photon counter and TCSPC module were both set at -80 mV, always lower than the voltages of the reflection peaks for different $I_b$, so the unwanted counts caused by the small reflection peak due to the impedance mismatch in the circuit can be avoided. Figure 3(b–d) show typical temporal histograms for SNSPDs with and without illumination ($n_{ph} = 10^5$ pps) at three different $I_b$ ($I_b = 0.95I_{sw0}$, $0.97I_{sw0}$, and $0.98I_{sw0}$). The blue histogram represents the distribution of dark counts in the time domain without illumination, the sum of which divided by the collecting time is the conventional DCR measured by the photon counter. The red lines obtained with illumination include two parts, the Gaussian peak and flat baseline, as noted in Fig. 3(b). The Gaussian peak denotes the photon counts, whereas the flat baseline denotes the dark counts. The baseline in Fig. 3(b) ($I_b = 0.95I_{sw0}$) appears similar to the blue histogram recorded without illumination. However, the baseline in Fig. 3(c) ($I_b = 0.97I_{sw0}$) is higher than the blue histogram, and the difference is apparent in Fig. 3(d) ($I_b = 0.98I_{sw0}$), which indicates that the DCR under illumination appears to be higher than that measured without illumination. However, the dark counts under illumination exhibit random behavior in the temporal histogram, which is different from that of the non-uniform dark counts (which exhibit either an extra peak or a wavy baseline) in previous reports [6, 18].

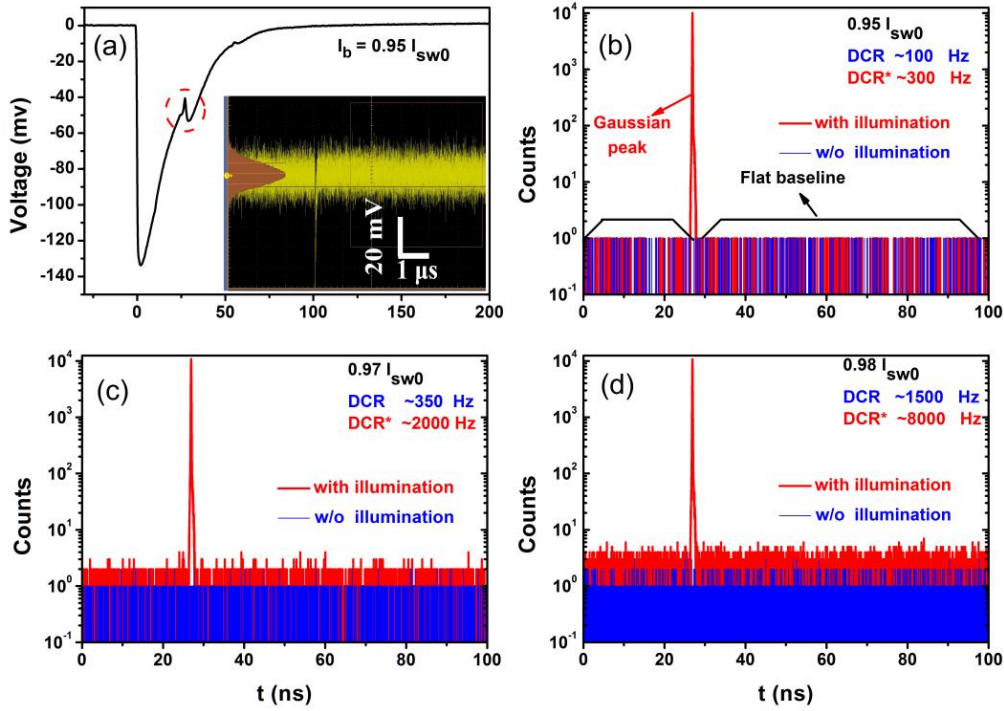

Fig. 3. (a) Response pulses of SNSPDs (averaged over 200 samples) at 0.95 $I_{sw0}$; the small abnormal peak, which is marked with the dashed circle, is due to impedance mismatch; the inset is the screen shot indicating the electronic noise floor. (b-d) semi logarithmic temporal histograms for counts from the SNSPD in logarithmic coordinates at 0.95 $I_{sw0}$, 0.97$I_{sw0}$, and 0.98$I_{sw0}$. $R_p$ is 10 MHz, $n_{ph}$ is $10^5$ pps, the time window is 100 ns, and the time bin is 4 ps, the total collecting time is 5 s. The red and blue lines represent the histograms with and without illumination, respectively. Note that many data points in (b-d) overlap due to the resolution limit on displaying or printing.

The TCSPC module provides a function to sum the counts for the entire histogram and the Gaussian peak denoting the photon count; therefore, we can accurately determine the SDE* and DCR* under illumination. The total counts $N_{tot}$ obtained from the TCSPC module are always the same as the counts clicked by the photon counter SR400. From the red lines with illumination, we may obtain SDE* = $N_p/(t\, n_{ph})$ and DCR* = $(N_{tot} - N_p)/t$, where $t$ is the collecting time and $N_p$ is the count in the Gaussian peak. Table 1 lists the SDE, DCR, SDE*, and DCR* data at three different $I_b$. All calculated SDE* have similar values when the $I_b$ is close to the $I_{sw}$, which indicates a saturated detection efficiency. However, the DCR* could not be neglected when $I_b = 0.97\ I_{sw0}$ or $0.98I_{sw0}$, which contributed absolute values of 1.5% and 6.4% to the SDE respectively. This finding explains why the SDE exhibits an abnormal increase and the increase in DCR* on SDE cannot be neglected.

Table 1. SDE, DCR, SDE*, and DCR* of SNSPD when $R_p$ = 10 MHz and $n_{ph}$ = $10^5$ pps

| $I_b/I_{sw0}$ | SDE (%) | SDE* (%) | DCR (Hz) | DCR* (Hz) |
|---|---|---|---|---|
| 0.95 | 77.3 | 77.1 | 92 | 288 |
| 0.97 | 78.1 | 76.6 | 350 | 1879 |
| 0.98 | 83.2 | 76.8 | 1400 | 7771 |

To further study the origin of the abnormal dark counts under illumination, $n_{ph}$ is changed from $10^5$ pps to $10^6$ pps, while maintaining the same repetition rate, i.e., $R_p$ = 10 MHz. The $I_b$ dependences of the SDE, SDE*, and DCR* for various $n_{ph}$ are shown in Fig. 4(a-b). The DCR as a function of $I_b$ is also shown in Fig. 4(b) for comparison. All DCR* curves are shifted toward smaller bias current with increasing $n_{ph}$. In other words, the dark counts increased with illumination at the same $I_b$. Figure 4(c) shows the enlarged details for Fig. 4(a), all SDE curves show abnormal increase when $I_b$ is approaching the $I_{sw}$. However, it is not so distinct when $n_{ph}$ is 0.5 MHz or 1MHz, since the DCR* of $10^4$ Hz contributes a smaller portion (2% and 1% respectively) to SDE. All SDE* curves exhibit a reasonable saturation behavior when the $I_b$ is close to $I_{sw}$ because all dark counts were deducted.

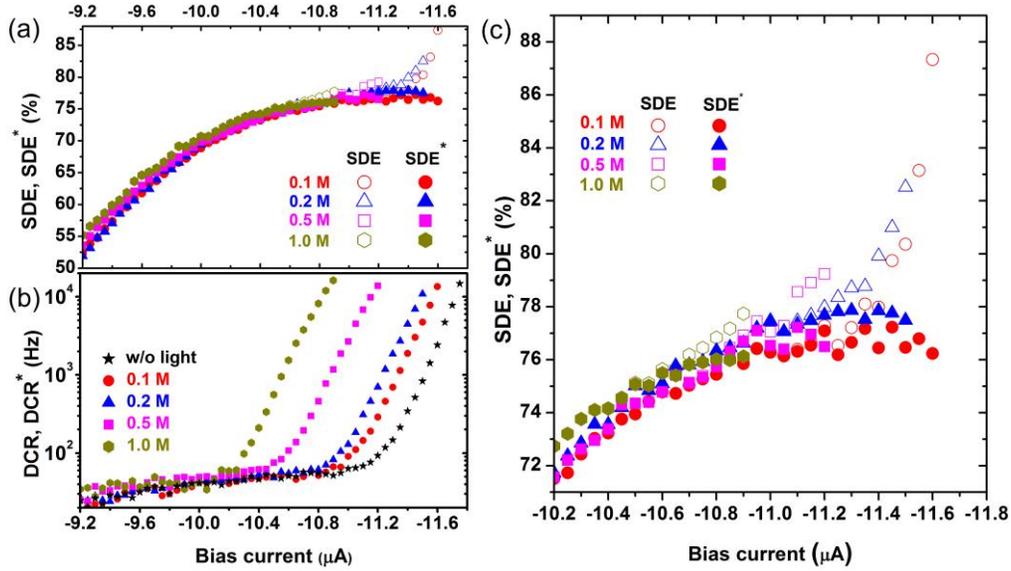

Fig. 4. Detection efficiencies and dark count rates of SNSPD as functions of $I_b$ when $n_{ph}$ is set to $10^5$ pps, $2 \times 10^5$ pps, $5 \times 10^5$ pps, and $10^6$ pps. (a) SDE and SDE*. (b) DCR and DCR*. The stars represent the conventional DCR without illumination. (c) Enlarged details of SDE and SDE*.

Figure 4 also indicates that the $I_{sw}$ under illumination appears to decrease with an increase in $n_{ph}$. To verify this behavior, the I–V characteristic of the SNSPD was examined and shown in Fig. 5(a). By tuning the voltage of the isolated voltage source (SIM928 from Stanford Research System) with a step of 2 mV, the current of SNSPD was swept with a step of 0.1 μA since an in-series resistor of 20 kΩ was adopted. The current and voltage values at each point in the $I$–$V$ curves were averaged values of 200 samples obtained by data acquisition card (NI 9215, National Instrument Corp). Indeed, we observed that the $I_{sw}$ decreased from −11.8 μA to −11.0 μA with an increase in $n_{ph}$ from 0 pps to $10^6$ pps. It is well known that $I_{sw}$ follows a certain distribution and varies from measurement to measurement [19]. To acquire the precise distribution of $I_{sw}$, we traced the $I$–$V$ curves for 1000 times under each illumination. A signal generator (WF1973, NF Corporation) with a sweeping rate of 1 kHz instead of the isolated voltage source was adopted to save the time for acquiring each trace. The statistics of $I_{sw}$ obtained from the $I$–$V$ curves gives normalized distributions of $I_{sw}$ shown in Fig. 5(b), which are roughly Gaussian. The peaks of the $I_{sw}$ distributions also decrease with the increase of $n_{ph}$, which is qualitatively consistent with the results from the $I$–$V$ curve in Fig. 4(a). In addition, since the DCR is exponentially proportional to $I_b/I_{sw}$ [20, 21], it is reasonable to observe a higher DCR when $I_{sw}$ is suppressed under illumination. We also noticed that $I_{sw}$ obtained from $I$-$V$ curves are always smaller than the peak value of the distribution. The possible reason is described as below. In the $I$-$V$ curves the switching current are determined by the highest supercurrent, which is actually the averaged values of 200 samples. If the switching happens during the sampling period, it will result in a smaller averaged switching current. Though we observed the suppression of $I_{sw}$ under illumination, the underlying mechanism is still unknown.

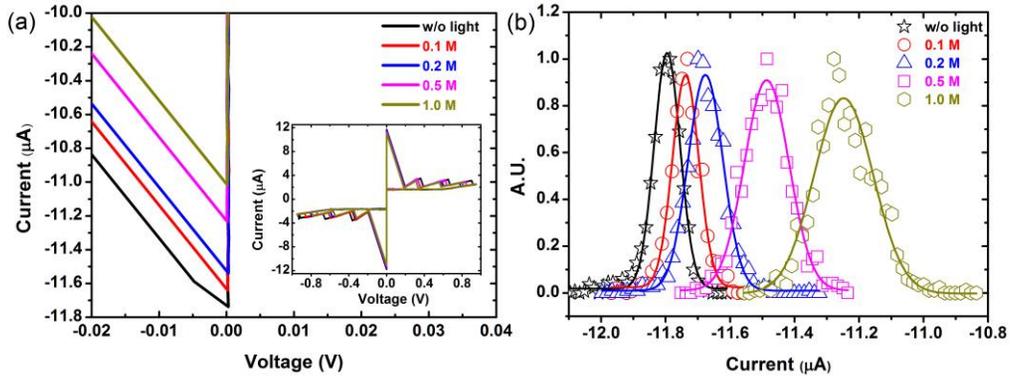

Fig. 5. $I$–$V$ characteristics and $I_{sw}$ distribution measurements of the SNSPD. (a) Enlarged $I$–$V$ curves indicating $I_{sw}$ of the SNSPD, from bottom to top corresponding to $n_{ph}$ of 1 Mpps, 0.5 Mpps, 0.2 Mpps, 0.1 Mpps, and w/o light respectively. The inset shows the full-scale $I$–$V$ curves. (b) $I_{sw}$ distribution of SNSPD under illumination with various $n_{ph}$, from left to right: w/o light (stars), 0.1 Mpps (circles), 0.2 Mpps (triangles), 0.5 Mpps (squares), and 1 Mpps (hexagons). The solid lines are the Gaussian fitting results.

Another experiment was performed to examine whether this phenomenon is related to the repetition rate of photons. We set $n_{ph} = 10^5$ pps and altered $R_p$ to 1 MHz, 5 MHz, 20 MHz, and 50 MHz successively. No difference was observed in the SDE* or DCR*, indicating that $n_{ph}$ results in an increase of dark counts regardless of the repetition rate. Actually, $n_{ph}$ was kept constant, thus the average time interval between two photon absorption events was also kept constant. As a result, it is reasonable to observe the same SDE* or DCR*. For the same reason, the above results will also occur for a CW laser though we cannot distinguish the photon counts and dark counts when the SNSPD is illuminated by a CW laser.

## 5. Discussions

In the above experiments, we demonstrated that the abnormal increase of the SDE is caused by the increase of the DCR when the $I_b$ is close to the $I_{sw}$. The underlying mechanism is the photon-induced suppression of the $I_{sw}$. A similar abnormal increase can also be observed in a previous report [13]. Indeed, the photon flux ($10^5$ pps, i.e., −109.9 dBm) adopted in this experiment is a very moderate value, which is comparable with or even smaller than the photon intensities selected in most experiments [9, 13, 22]. Additionally, from an application viewpoint, the value we selected is reasonably small. For example, in a short-distance and high-speed QKD, a raw key rate of 100 kbps corresponds to an arrival $n_{ph}$ of $10^5$ pps, assuming that SDE = 100%. In this case, the DCR of the detector should be re-evaluated for QKD rather than using the conventional DCR measured without illumination. If QKD needs a detector DCR of <100 Hz and $I_b$ is set at 12.0 µA, as suggested by Fig. 1, the real DCR* reaches 288 Hz, which may result in an unexpected larger quantum error bit rate. One needs to bias the detector at a smaller current such that the DCR* is consistent with the DCR. When $n_{ph}$ is high (e.g., $10^6$ pps or higher), the photon counts are much larger than the dark counts even if the dark counts are effectively increased, as demonstrated by the dark yellow hexagons in Fig. 4(c). This phenomenon explains why the abnormal increase in SDE was not observed in most experiments.

A high-performance SNSPD with an SDE over 75% at a DCR of 100 Hz was fabricated. An abnormal increase in the SDE was observed when the $I_b$ was close to the $I_{sw}$, with photon flux being as low as $10^5$ pps. Temporal analysis of the detection counts using a TCSPC module indicated that illumination results in an increase of the DCR and in an abnormal increase of the SDE. The DCR increase is related to the suppression of the $I_{sw}$ of the nanowire under illumination. From an application viewpoint, it is necessary to clarify that the DCR of an SNSPD under illumination at a high $I_b$ can be different from the DCR measured without illumination.

**Acknowledgments**

This study was partially supported by the National Natural Science Foundation of China (Grant 61401441 & 61401443), 973 Program (Grant 2011CBA00202), and partially by the "Strategic Priority Research Program (B)" of the Chinese Academy of Sciences (Grant XDB04020100 & XDB04020100).